\documentclass[%
 aip,
 twocolumn,
 jmp,%
 amsmath,amssymb,
reprint,%
]{revtex4-1}

\usepackage{graphicx}
\usepackage{dcolumn}
\usepackage{bm}
\usepackage{gensymb}
\usepackage{epsf}
\usepackage{epsfig}

\newcommand{\dg}{~\mathrm{deg} }
\newcommand{\um}{~\mu\mathrm{m} }
\newcommand{\mm}{~\mathrm{mm} }
\newcommand{\cm}{~\mathrm{cm} }

\begin{document}

\preprint{AIP/123-QED}

\title[]{Multiangle static and dynamic light scattering in the intermediate scattering angle range}
\author{E. Tamborini}
\email{elisa.tamborini@univ-montp2.fr} \affiliation{Universit\'{e} Montpellier
2, Laboratoire Charles Coulomb UMR 5221, F-34095, Montpellier,
France} \affiliation{CNRS, Laboratoire Charles Coulomb UMR 5221,
F-34095, Montpellier, France\\}

\author{L. Cipelletti}
\affiliation{Universit\'{e} Montpellier 2, Laboratoire Charles
Coulomb UMR 5221, F-34095, Montpellier, France} \affiliation{CNRS,
Laboratoire Charles Coulomb UMR 5221, F-34095, Montpellier,
France\\}

\date{\today}

\begin{abstract}
We describe a light scattering apparatus based on a novel optical scheme covering the scattering angle range $0.5\dg \le \theta \le 25\dg$, an intermediate regime at the frontier between wide angle and small angle setups that is difficult to access by existing instruments. Our apparatus uses standard, readily available optomechanical components. Thanks to the use of a charge-coupled device detector, both static and dynamic light scattering can be performed simultaneously at several scattering angles. We demonstrate the capabilities of our apparatus by measuring the scattering profile of a variety of samples and the Brownian dynamics of a dilute colloidal suspension.
\end{abstract}

\maketitle

\section{Introduction}
\label{sec:introduction}

Light scattering methods are a powerful tool for investigating the structure and the dynamics of soft and biological matter. The typical space and time scales that are probed range from tens of nanometers to tens of microns, and from a fraction of a microsecond to several hours, respectively. Typical applications include particle sizing, the characterization of aggregation phenomena, the determination of interparticle interactions, the investigation of the structure and the relaxation dynamics of complex fluids. The diversity of systems that can be studied makes the technique appealing for both academic laboratories and industry: suspensions of colloidal particles or emulsions, solutions of polymers and proteins, foams, and surfactant phases are but a few examples of such systems.

Scattering methods may be divided in two classes: static and dynamic light scattering. In static light scattering (SLS)~\cite{Kerker}, one probes the structure of a sample by measuring the time-averaged scattered intensity as a function of the scattering angle. By contrast, dynamic light scattering (DLS)\cite{Berne1976} focuses on the temporal fluctuations of the scattered intensity in order to extract valuable information on the sample dynamics. In both cases, the sample is typically illuminated by a laser beam, while photodiodes, avalanche photodiodes, photomultiplier tubes or CCD or CMOS cameras are used as a detector.

A wide variety of light scattering apparatuses have been developed, many of which are commercially available. The design of a setup is in general optimized according to the range of scattering angles to be covered and the kind of measurements to be performed (static or dynamic). Wide-angle light scattering (WALS) setups often provide both SLS and DLS measurements, covering approximately the range $10\dg \le \theta < 180\dg$, corresponding to length scales of order $2\pi q^{-1}$ ranging from $0.2\um$ to $2.3\um$. Here $q = 4\pi n \lambda^{-1} \sin (\theta/2)$ is the scattering vector, with $n$ the solvent refractive index, $\lambda$ the in-vacuo laser wavelength and $\theta$ the scattering angle. However, it should be noted that in practice it is difficult to obtain reliable data below $ \theta = 20\dg$, because the cylindrical sample geometry that it is usually adopted in WALS limits the quality of the optical interfaces. Small-angle and ultra-small-angle light scattering setups (SALS and USALS, respectively) are specialized for measurements close to the forward direction, with scattering angles covering up to two decades, up to approximately $10\dg$ for SALS and a few degrees for USALS. The corresponding length scales vary between a few microns and tens or even hundreds of microns. While some of these setups use dedicated photodiode detectors optimized for SLS but unfit for DLS~\cite{Ferri1997,glatter98,Ferri2001,Ferri2002}, the adoption of a CCD or CMOS detector is increasingly popular, since it allows for both static and dynamic light scattering to be performed~\cite{WongRSI1993,Cipelletti1999}, as first shown by the pioneering work of Wong and Wiltzius~\cite{WongRSI1993}.

It should be noted that there is little if any overlap between the range of scattering angles of typical WALS and SALS or USALS setups: this leaves uncovered a crucial angular range corresponding to probed length scales on the order of a micron, the characteristic size of many colloidal objects of industrial and fundamental science interest. Additionally, absolute intensity measurements are notoriously difficult in light scattering, especially for SALS or USALS, thus making it difficult to merge data from different setups on the same scale. To circumvent this problem, a few apparatuses covering the ``mid angle'' range (mid-angle light ccattering, MALS) have been proposed in the past. The apparatus described in Ref.~\cite{glatter98} covers an impressive angular range ($2\deg - 60\deg$), using a custom made cell and dedicated photodiode arrays and electronics. Ferri and coworkers report a setup covering scattering angles up to $15\deg$~\cite{Ferri2001,Ferri2002}, using a commercially available cell but again a custom photodiode array with dedicated electronics. A different approach is used in the commercially available apparatus of Ref.~\cite{Webb2000}, which covers scattering angles from a fraction of degree up to about $40\deg$ by varying the propagation direction of the incident beam. Accordingly, several distinct measurements are required to sample the full angular range, each measurement covering about $5\deg$. Chou and Hong~\cite{Chou2003} report a CCD-based setup in the range $2\deg < \theta < 25\deg$ using a scheme originally proposed by Ferri for SALS~\cite{Ferri1997}. Unfortunately, however, no detailed description and characterization of the apparatus performances are provided. Finally, one may take advantage of the good-quality optics of modern microscopes to build an apparatus that combines imaging with low- or mid-angle scattering. Kaplan \textit{et al.}~\cite{kaplan99} report a DLS apparatus based on an inverted microscope that covers scattering angles between $20.6\deg$ and $55.1\deg$, while Celli \textit{et al.}~\cite{Celli05} demonstrate DLS in a upright microscope for $\theta \le 12\deg$. In both cases, measurements are performed at one single angle at a time. In Ref.~\cite{valentine01}, a microscope- and CCD-based static light scattering instrument is presented, covering the range $0.9 \um \le q \le 18\um$, corresponding approximately to $3.3\deg \le \theta \le 70\deg$.

With the exception of the setups of Refs. ~\cite{Chou2003,valentine01} that could in principle be extended to DLS,
these apparatuses are unfit for simultaneous static and dynamic light scattering at multiple angles, either because the angular range is sampled sequentially, as in Refs.~\cite{Webb2000,kaplan99,Celli05}, or because photodiodes that cover a very large number of speckles are used, as in Refs.~\cite{glatter98,Ferri2001,Ferri2002}, a design not appropriate for DLS~\cite{Berne1976}. It is worth noting that photodiode arrays are typically larger than CCD or CMOS detectors: this makes it difficult to transpose the optical layout of Refs.~\cite{glatter98,Ferri2001,Ferri2002} to a CCD-based apparatus and would impose extra constraints on the cell dimensions, in particular its thickness $L$, as we shall discuss it in the following. Additionally, photodiode arrays can easily accommodate a small hole to let through the transmitted beam, a crucial requirement for small angle measurements, while this is not possible for a CCD or CMOS detector. Thus, existing designs for MALS cannot be easily generalized to an apparatus for both SLS and DLS.

In this paper, we present a CCD-based mid-angle light scattering apparatus that covers the intermediate range $0.5\dg \le \theta \le 25\dg$ thanks to a novel optical layout. Our MALS setup allows for simultaneous measurements at all accessible scattering angles. Both SLS and DLS measurements can be performed and no stringent limitations on the sample thickness are imposed. Finally, the setup uses only standard components that are readily available off the shelf. The rest of the paper is organized as follows: To better understand the difficulties in extending the angular range of SALS or USALS setups to larger angles, we review some of the most popular optical layouts in Sec.~\ref{sec:layout}. We then present our new MALS apparatus in Sec.~\ref{sec:MALS}, before describing its angular calibration in Sec.\ref{sec:calibration}. A series of tests of the apparatus' performances for both SLS and DLS are presented in Sec.~\ref{sec:test}, before the concluding remarks of Sec.~\ref{sec:conclusions}.

\section{Popular optical layouts for SALS and USALS}
\label{sec:layout}
\begin{figure}
\includegraphics[scale=0.5]{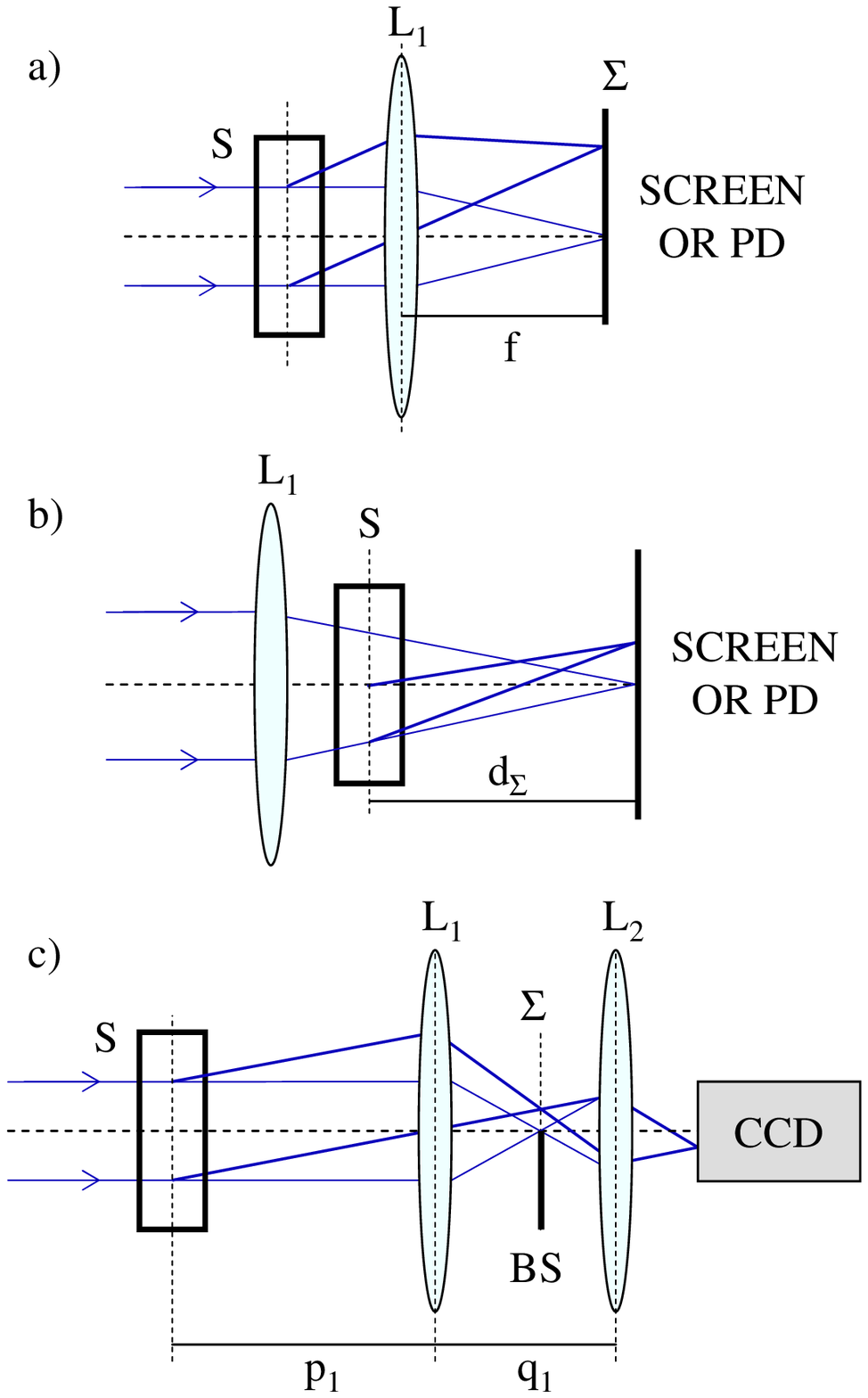}
\caption{\label{fig:intro}Scheme of common SALS and USALS setups. S: sample; $L_1$, $L_2$: lenses; PD: photodiode array; $\Sigma$: focal plane of $L_1$, BS: beam stop.}
\end{figure}

Most MALS setups are based on the same optical layout as that for SALS and USALS, examples of which are shown  in Fig.~\ref{fig:intro}. In the top scheme, Fig.~\ref{fig:intro}a, adopted in Refs.~\cite{M.CarpinetiF.FerriM.GiglioE.Paganini1990,glatter98,Bassini1998,Webb2000}, the scattered intensity is measured in the focal plane $\Sigma$ of a so-called Fourier lens of focal length $f$. Neglecting refractions at the solvent-cell and cell-air interfaces, a point on $\Sigma$ at a distance $r$ from the optical axis corresponds to a scattering angle $\theta = \arctan(rf^{-1})$. In order to avoid artifacts due to light leaking from the intense transmitted beam, a hole may be drilled in the detector to let the transmitted beam pass through $\Sigma$~\cite{M.CarpinetiF.FerriM.GiglioE.Paganini1990,glatter98,Bassini1998,Ferri2001,Ferri2002}. Alternatively, a screen may be placed in the plane $\Sigma$ and a CCD camera may be used to record the intensity distribution on the screen. In both cases, this geometry prevents a CCD camera to be placed directly in the plane $\Sigma$, making DLS measurements impossible. In addition, detectors and lenses used in this configuration must often be custom-made to efficiently remove the transmitted beam and to limit aberrations associated with large scattering angles. The scheme of Fig.~\ref{fig:intro}b, described in Refs.~\cite{Ferri2001,Ferri2002}, is equivalent to the top one, provided that one replaces $f$ by the sample-detector distance $d_{\Sigma}$ in the calculation of $\theta$~\cite{goodmanfourieroptics}. The advantage is that large scattering angles may be attained simply by reducing $d_{\Sigma}$, without changing $f$ and with no stringent requirements on the numerical aperture of the Fourier lens. However, the sample thickness $L$ must satisfy $L<< d_{\Sigma}$, since the mapping between $r$ and $\theta$ depends on the distance of a scatterer from $\Sigma$. When using a CCD detector, typical values of $d_{\Sigma}$ are a few cm at most, limiting $L$ to about 1 mm. Additionally, the remarks on removing the transmitted beam discussed for the top scheme apply also here. Thus, the middle scheme is limited to SLS measurement on thin samples. The scheme of Fig.~\ref{fig:intro}c was proposed by Ferri for a CCD-based SALS apparatus~\cite{Ferri1997} and extended to larger angles by Chou and Hong~\cite{Chou2003} and to DLS by Cipelletti and Weitz~\cite{Cipelletti1999}. It is similar to the top scheme, except that a beam block is placed in $\Sigma$ and a second lens is used to image the intensity distribution in the Fourier plane onto a CCD. This scheme can not be easily extended to MALS because the numerical aperture of commercially available lenses limits in practice the maximum scattering angle to about $15\dg$, beyond which vignetting sets in.

\section{The MALS setup}
\label{sec:MALS}

\begin{figure*}
\includegraphics[scale=0.9]{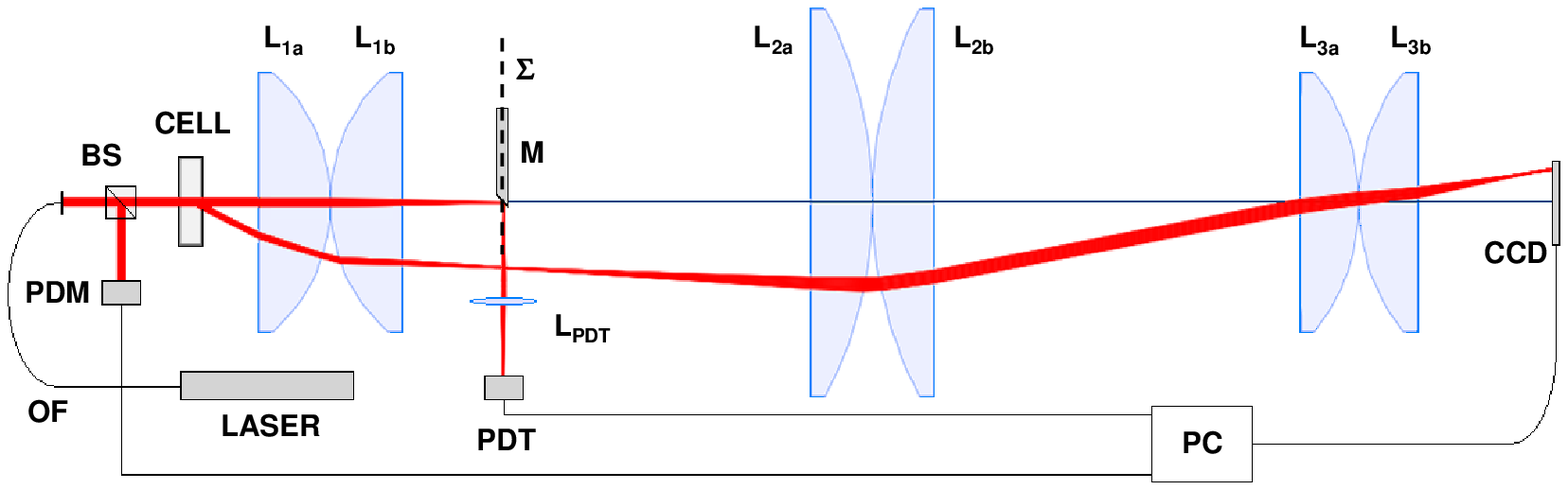}
\caption{\label{fig:setup}The Mid-Angle Light Scattering apparatus described
in this paper. The two
pairs of lenses $L_{1a}$, $L_{1b}$ and $L_{2a}$, $L_{2b}$ are confocal
and act as an (inverted) telescope. The beam stop M is placed in
their common focal plane $\Sigma$ to remove the transmitted beam. The pair of lenses $L_{3a}$, $L_{3b}$
acts as a Fourier lens and the CCD detector is placed in its back focal plane. OF: optical fiber; BS: beam splitter; PDM, PDT: monitor and transmitted photodiode, respectively; PC: personal computer. The optical elements are drawn in scale, the cell-CCD distance is $28.8\cm$.}
\end{figure*}

A scheme of our MALS setup is shown in Fig.~\ref{fig:setup}. Details of the various optical and opto-mechanical components and of their position are provided in the Supplementary Information. In order to increase the numerical aperture of the various elements, we use pairs of identical lenses placed in contact, for which the effective focal length is half of that of each lens, thereby doubling the numerical aperture. The apparatus maps the cone of light scattered at a given angle $\theta$ onto a ring of CCD pixels centered around the transmitted beam position. Functionally, it may be divided into two parts. The first part is formed by the two pairs of lenses $L_{1a},L_{1b}$ and $L_{2a},L_{2b}$, which are confocal and act as an (inverted) telescope. Their purpose is two-fold: on the one hand, they demagnify the angle of the light scattered by the sample, thus mapping the MALS range to a more manageable, lower-angle range. On the other hand, a beam stop can be conveniently placed in the focal plane $\Sigma$ common to both pairs, thereby removing the transmitted beam. The second part comprises the pair of lenses $L_{3a},L_{3b}$ that act as a Fourier lens, in whose back focal plane a CCD detector is placed. The focal length of the various lenses is $f_{1a}=f_{1b} =
62.9\mm$, $f_{2a}=f_{2b} = 150\mm$ and $f_{3a}=f_{3b} = 75.6\mm$. The resulting demagnification ratio of the inverted telescope is $f_{2a}/f_{1a} \approx  2.4:1$. In order to cover the MALS range, we use plano-convex lenses and work well beyond paraxial propagation; therefore, the thin lens approximation does not apply. We use a ray-tracing software (Optical Ray Tracer 2.8~\cite{WebRayTracing}) to test the performances of the real lenses and to fine tune the position of the sample and that of the third pair of lenses. To avoid vignetting, the sample must be close enough to $L_{1a}$, while a minimum distance of about 5 mm is imposed by the thickness of the cell wall and the sample and lens holder. Using the ray-tracing software, we determine the optimum sample-$L_{1a}$ and $L_{3b}$-CCD distances to be $17\mm$ and $30\mm$, respectively. With this layout, the maximum relative spread of the spot formed on the CCD plane by light scattered at the same $\theta$ is $1.7\%$, for $\theta = 25\dg$ and a $10\mm$ thick sample.

The light source is an intensity-stabilized, linearly polarized He-Ne
laser (117A by Spectra-Physics) that operates at a wavelength $\lambda = 632.8 ~\mathrm{nm}$, and at a
power of $1.0~\mathrm{mW}$. The
laser beam is coupled to a polarization-maintaining single-mode
fiber optics (LPC-04-633-4/125-P-0.9-4.5AS-60-A3A-3A-2 by OZ Optics), through which only the $\mathrm{TEM}_{00}$ mode can propagate. A
set screw is used to partially block the beam before coupling it to the
fiber, so that the beam power can be attenuated as required. The
beam exiting the fiber is collimated to a diameter of
$0.9\mm$ and it is directed on a beam splitter (BS) that partially
forwards it onto a photodiode (PDM), which monitors any fluctuations
in the incident power for further data normalization. The transmitted component impinges onto the
sample, which is typically contained in a parallelepiped cell of
thickness 2, 5 or $10\mm$ (Hellma). A small mirror M is placed in the focal plane $\Sigma$ to intercept the transmitted beam and direct it to a
second photodiode (PDT), allowing for the measurement of the sample
transmission and preventing unscattered light from reaching the CCD
sensor. Following Ref.~\cite{Ferri1997}, the mirror is obtained by cutting at $45\dg$ and polishing a drill bit of thickness $300\um$. To minimize light reflections, all lenses are antireflection
coated and are tilted by a few degrees with respect to the optical
axis. The same precaution is taken for the
scattering cell. The cell is mounted on a custom-designed
holder that allows one to vary both its height with respect to
the optical axis and its distance from $L_{1a}$. The cell can be removed from the holder and replaced exactly at the same position, i.e. for taking an optical background as described later. The CCD camera (Pulnix TM-1300) has a
pixel matrix of $1285 \times 1029$ pixels, $6.7 ~\mathrm{\mu m}$ in size. A frame grabber (Meteor II Digital PCI by Matrox) is used to control the camera and transfer the images to a PC. The whole instrument is placed under a box made of high density expanded polystyrene panels, to minimize temperature fluctuations and to protect the setup from ambient light.

\section{Data acquisition and processing}
\label{sec:processing}

We use a custom C++ code to process the CCD images in real time, in order to perform both static and dynamic light scattering measurements.

\subsection{Static Light Scattering}
Data acquisition and processing for SLS involves four steps: i) acquisition of the electrical background, ii) acquisition of the optical background, iii) acquisition of the scattered intensity pattern, iv) applying to iii) corrections for the contribution of i), ii) and geometrical factors in order to obtain the scattered intensity profile $I(q)$. During step i), the laser beam is blocked so as to measure the contribution of electrical noise to the data. The signals of the monitor and transmitted photodiodes, $M^{\text{EB}}$ and $T^{\text{EB}}$, are averaged over 100 readings and recorded. Similarly, a series of CCD images taken at various exposure times $t_{\text{exp}}$ are recorded, about 20 images being averaged for each $t_{\text{exp}}$. The images are processed in order to calculate the dark background averaged over annuli of pixels centered around the transmitted beam position: $S^{\text{EB}}(q,t_{\text{exp}}) = \left < S^{\text{EB}}_p(t_{\text{exp}})\right >_{p \in A_q}$, where $S^{\text{EB}}_p(t_{\text{exp}})$ is the CCD electrical background signal for the $p-$th pixel at a given exposure time and $\left< \cdot \cdot \cdot \right>_{p \in A_q}$ indicates an average over the annulus of pixels associated with a given $q$ vector. Pixels that are covered by the beam block are software-masked and excluded from the analysis. The purpose of step ii) is to measure any contributions to the scattered light due to flare, originating from dust or imperfections on the cell walls and the lenses. To this end, the cell is filled with the solvent alone and the optical background signals for the photodiodes and the CCD are recorded in analogy to step i): $M^{\text{OB}}$, $T^{\text{OB}}$ and $S^{\text{OB}}(q,t_{\text{exp}})$. Before step iii), the cell is emptied from the solvent and refilled with the sample. For the optical background correction to be effective, the cell should be kept exactly in the same position, which can be achieved by emptying and refilling it \textit{in situ}. When this is not possible (e.g. for samples that need to undergo a controlled thermal history, as in Ref.~\cite{tamborini12}), we use a custom-designed holder that allows the cell to be removed and replaced with a positioning tolerance of about $10 \um$. During step iii), the intensity scattered by the sample at time $t$ is recorded, together with the corresponding photodiode signals:  $S(q,t_{\text{exp}},t)$, $M(t)$ and $T(t)$. In order to reduce noise, a few images for each $t_{\text{exp}}$ may be averaged. Depending on the number of averages and that of the exposure times, one full acquisition may require as little as 0.1 sec. Note that the use of different exposure times is often mandatory, given the limited dynamic range of CCD detectors and the steep variation of $I(q)$ typically observed over the MALS range. Data stored in steps i) to iii) are processed in real time in order to calculate the $q$ dependent scattering intensity according to
\begin{widetext}
\begin{equation}
I(q,t)=\frac{1}{L N(q)t_{\text{exp}}^*(M(t) - M^{\text{EB}})}
\left\{S(q,t_{\text{exp}}^*,t)-S^{\text{EB}}(q,t_{\text{exp}}^*) - \frac{T(t)-T^{\text{EB}}}{T^{\text{OB}}-T^{\text{EB}}} \left[
S^{\text{OB}}(q,t_{\text{exp}}^*)-S^{\text{EB}}(q,t_{\text{exp}}^*)\right]\right\} \,.
\label{eq:finalInt}
\end{equation}
\end{widetext}
In the above equation, all optical signals are corrected for the corresponding electrical background and normalized with respect to the incident power as measured by the monitor photodiode. Before subtracting the optical background (the term between square brackets in the r.h.s. of Eq.~(\ref{eq:finalInt})), we correct it for the sample transmission, $(T(t)-T^{\text{EB}})/(T^{\text{OB}}-T^{\text{EB}})$, because the flare contribution is decreased when part of the incident beam is scattered by the sample. In Eq.~(\ref{eq:finalInt}), $t_{\text{exp}}^*$ is the best exposure time for a given $q$, allowing for a good signal to noise ratio while preventing pixel saturation. We find that for a 8-bit CCD $t_{\text{exp}}^*$ should be chosen as the largest exposure time such that $S(q,t_{\text{exp}}^*)\lesssim 40$ grey levels. For larger values, the number of saturated pixels becomes non-negligible, while for smaller values data are affected by digitalization noise. Note that the intensity is normalized with respect to cell thickness $L$ and exposure time $t_{\text{exp}}^*$, and by the term $N(q)$, which accounts for the average value of the solid angle associated to each pixel of a given annulus, as well as for the dipole term $\sin^2\psi_p$ that decreases the scattered intensity out of the scattering plane~\cite{Berne1976}, where $\psi_p$ is the angle between the polarization of the incident light and the propagation direction of the scattered light that reaches the $p$-th pixel. More specifically, $N(q) =  <\Delta\Omega_p\sin^2\psi_p)>_{p\in A_q}$, with $\Delta\Omega_p=\sin\theta_p\Delta\theta_p\Delta\varphi_p$ the solid angle, $\sin^2\psi_p= 1-\sin^2\varphi_p\sin^2\theta_p$ the dipole factor, $\theta_p$ the scattering (polar) angle, and $\varphi_p$ the azimuthal angle. For a pixel $p$ with coordinates $(x,y)$ with respect to the transmitted beam position one has
\begin{subequations}
\begin{eqnarray}  \Delta\theta_p=\dfrac{1}{r}(x\Delta x + y\Delta y)\dfrac{d\theta_p}{dr} \\
\label{eq:deltatheta}
\Delta\varphi_p=\dfrac{d\varphi_p}{dx}\Delta x+ \dfrac{d\varphi_p}{dy}\Delta y \,,
\label{eq:deltaphi}
\end{eqnarray}
\end{subequations}
with $r = \sqrt{x^2 + y^2}$, $\varphi_p=\arctan(y/x)$, and $\Delta x$ and $\Delta y$ the pixel dimensions. The functional form relating $\theta_p$ to $r$ and from which the term $\dfrac{d\theta_p}{dr}$ can be calculated will be discussed in Sec.~\ref{sec:calibration}.

\subsection{Dynamic Light Scattering}
The same four steps discussed in the previous subsection are also required for DLS measurements, although of course the data processing in step iv) is different. We use a software multi-tau correlator~\cite{multitau}, whose details are discussed in Ref.~\cite{Cipelletti1999}, together with the algorithm for correcting for the electrical and optical background. Here, we simply recall that the correlator calculates in real time the time autocorrelation function of the CCD signal, $\left<\left<S_p(t)S_p(t+\tau)\right>_{p \in A_q}\right>_t$. After correction for the electrical and optical backgrounds and proper normalization, using the Siegert relation~\cite{Berne1976} one obtains the field autocorrelation function
\begin{equation}g_1(q,\tau) = \frac{\left<E(q,t)E^*(q,t+\tau)\right>_t}{\left<I(q,t)\right>_t} \,,
\end{equation}
which is directly related to the sample dynamics~\cite{Berne1976}.

\section{Angular calibration}
\label{sec:calibration}

\begin{figure}
\includegraphics[scale=1.8]{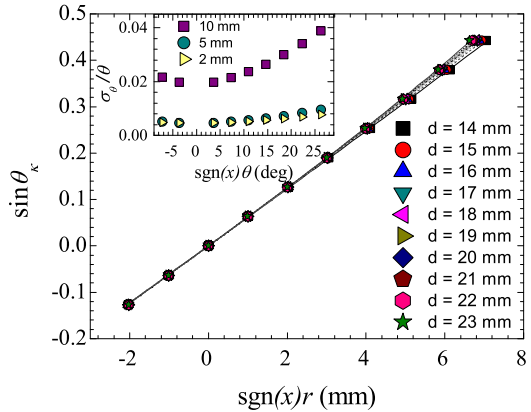}
\caption{\label{fig:diffGrat} MALS angular calibration
performed using a diffraction ruling with $m=1000~\text{lines} \cm^{-1}$.
In the main plot the sine of the angular position of the the maxima of the
diffracted beam is shown as a function of the maxima position on the CCD plane, for distances $d$ between the diffraction grating and lens $L_{1a}$ between 14 and 23 mm. The symbols are the data, the lines are fits according to Eq. ~(\ref{eq:sintheta}). The inset shows the loss of
angular resolution, as defined in the text, due to a sample thickness of $2\mm$,$5\mm$ and $10\mm$.}
\end{figure}

With the optical layout of Fig.~\ref{fig:setup} and in the thin lens and paraxial approximations, the relationship between the scattering angle $\theta$ and the distance $r$ from the optical axis of a CCD pixel is
\begin{equation}
\sin\theta = r(r^2 + f_{\text{eff}}^2)^{-0.5}\,,
\label{eq:feffective}
\end{equation}
where $f_{\text{eff}} = f_{3a}f_{2a}/ (2f_{1a}) \approx 15.9 \mm$ is the effective focal length of the Fourier lens, which accounts for both the use of pairs of lenses and the demagnification of the inverted telescope. Note that for the moment we have neglected refraction, assuming that the scatterers are in air. Since deviations are to be expected with respect to this simple formula, we perform an angular calibration of the setup using a diffraction ruling with $m=1000$ $\text{lines}\cm^{-1}$. The diffraction grating is positioned perpendicular to the incident beam, so that the maxima of the diffracted beam occur at angles such that $\sin\theta_{k} = mk\lambda$, $k = 0,\pm1,\pm2...$. We vary the distance $d$ between the diffraction grating and $L_{1a}$ between 14 and 23 mm in steps of 1 mm and for each $d$ we measure the position $r$ of the maxima on the CCD sensor. The results are shown in Fig.~\ref{fig:diffGrat}. The highest order observable is $k=7$, corresponding to $\theta_{k}= 26.3\dg$, which we achieve by offsetting the CCD with respect to the optical axis, such that the transmitted beam impinges near one corner of the detector, rather than in its center. The distance $d$ for which the maxima appear to be sharper is found to be $18\mm$, close to $d= 17\mm$ as evaluated using the ray-tracing software. Thus, thick samples are always positioned such that the cell center is at 18 mm from the first lens. Figure~\ref{fig:diffGrat} shows that $\sin \theta_k$ varies approximately linearly with $r$, as expected from Eq.~(\ref{eq:feffective}) given that $f_{\text{eff}}^2 >> r^2$. However, deviations are observed at large $r$; more importantly, these deviations depend on $d$, implying that lens aberrations limit the angular resolution of the setup for samples of finite thickness. To quantify these effects, we fit for each $d$ the data of Fig.~\ref{fig:diffGrat} using the empirical relation
\begin{equation}
\sin\theta_k=ar+
\dfrac{1}{b-r}-\dfrac{1}{b} \,,
\label{eq:sintheta}
\end{equation}
with $a$ and $b$ $d$-dependent fitting parameters and where $r$ is measured in mm. As seen in Fig.~\ref{fig:diffGrat}, this expression is in excellent agreement with the data. For samples of finite thickness and for a solvent of refractive index $n$, we define the scattering angle associated to a pixel $p$ laying at a distance $r$ from the optical axis as
\begin{equation} \theta_p=\arcsin{\left [\frac{1}{n}\left(\bar{a}r+
\dfrac{1}{\bar{b}-r}-\dfrac{1}{\bar{b}}\right)\right ]} \,,
\label{eq:teta}
\end{equation}
where the $n^{-1}$ factor accounts for refraction at the solvent-wall and wall-air interfaces and where $\bar{a}$ and $\bar{b}$ are the average of the fitting parameters $a$ and $b$ over the range of $d$ corresponding to the sample thickness. For a 10 mm-thick cell, we find $\bar{a}^{-1} = 17.27 \mm$, about $8.6\%$ larger than $f_{\text{eff}}$ as calculated neglecting aberrations. The $q$ vector associated with an annular set of pixels is obtained by averaging the corresponding value for each pixel of the annulus:
\begin{equation}
q= \dfrac{4\pi n}{\lambda}\left < \sin \left(\frac{\theta_p}{2}\right) \right >_{p \in A_q} \,
\label{eq:qexp}
\end{equation}
with $\theta_p$ as obtained from Eq.~(\ref{eq:teta}).

The inset of Fig.~\ref{fig:diffGrat} quantifies the loss of angular resolution due to sample thickness by showing the relative standard deviation of $\theta$, $\sigma_{\theta}/\theta$, associated with the standard deviation of the coefficients $a$ and $b$ over the range of $d$ corresponding to representative values of the cell thickness $L$. For $L \le 5\mm$, the relative angular spread is well below $1\%$ over the full range covered by the setup, while for a 10 mm-thick cell, the loss of angular resolution is about $4\%$ at the largest $\theta$.

\section{Tests of the apparatus performances}
\label{sec:test}

We perform a series of experiments to validate the angular calibration determined in Sec.~\ref{sec:calibration}, determine the angular response of the apparatus and to test its performances for DLS.

\subsection{Static light scattering}

Static light scattering tests were performed on representative 2-dimensional and 3-dimensional samples. For all samples, $I(q)$ was measured for pixel annuli whose average radius ranged from $29 ~\mathrm{pixels}$ to $1279 ~\mathrm{pixels}$, for a total of 126 values equally spaced in a linear scale. The thickness of the annuli is
$1 ~\mathrm{pixel}$. To cope with the large variation of the scattered intensity over the range covered by the instrument,
eighteen exposure times have been used, from $0.08~\mathrm{ms}$ to $80~\mathrm{ms}$, with a scaling factor of 1.5 between one exposure time and the next one.

\begin{figure}
\includegraphics[scale=1.8]{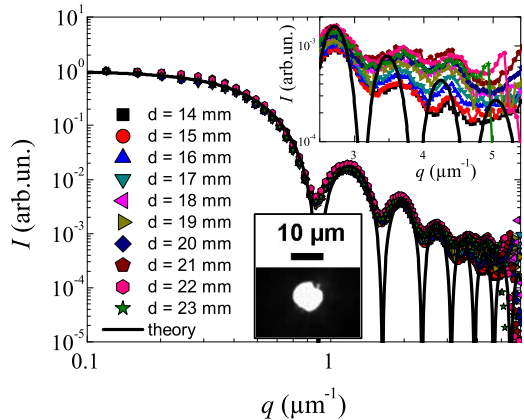}
\caption{\label{fig:ph} Diffraction pattern of a pinhole with a nominal diameter of $10\um$
measured for ten values of the sample-$L_{1a}$ distance $d$.
In the inset, the discrepancy in the $q$-calibration and the intensity level
for the ten distances at high $q$ vectors are visible.
An image of the pinhole taken with an optical microscope is also shown.}
\end{figure}

To verify the $q$ calibration and to check for the angular response of the apparatus, we measure the diffraction pattern of a
pinhole with a nominal diameter of $10 ~\mathrm{\mu m}$. Inspection under an optical microscope reveals a real diameter of $(8.66 \pm 0.24 ~\mathrm{\mu m})$ and some deviations from a perfectly circular shape (inset of Fig.~\ref{fig:ph}). The
diffraction pattern of the pinhole was measured by placing it at ten
distances between $14\mm$ and $23\mm$ from $L_{1a}$ and centering it
onto the optical axis. The beam stop was removed. The ten intensity
distributions are shown in Fig.~\ref{fig:ph} (symbols, data rescaled so that $I(q \rightarrow0)$ = 1), together with the Airy function predicted by the Fraunhofer diffraction theory (line).
All curves agree well with the theory up to $q \approx 2.5\um^{-1}$. Beyond this value, the
$q$ calibration is still good up to $q = 5\um^{-1}$, but deviations appear with respect to the theory. While this discrepancy may be due in part to the shape defects of the pinhole, we observe that the intensity
level is systematically lower than the theoretical one for
$d \le 16\mm$ and systematically higher for $d \ge 20\mm$, as better seen in the inset of Fig.~\ref{fig:ph}.

\begin{figure}
\includegraphics[scale=1.8]{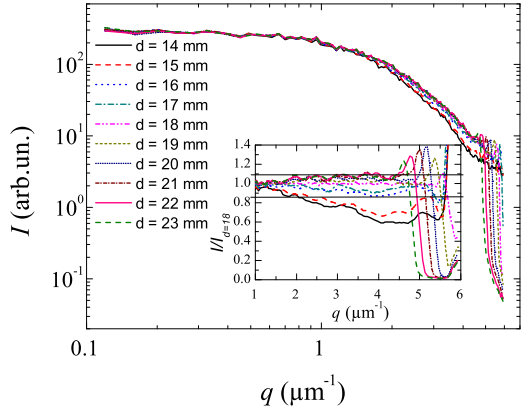}
\caption{\label{fig:glass}Scattering from a ground glass positioned at
distances $d$ varying between $14\mm$ and $23\mm$. The inset shows
the ratio of the measured intensity with respect to that recorded for $d=18\mm$, corresponding to the optimum sample-$L_{1a}$ distance.}
\end{figure}

To better quantify the $d$-dependent response of the setup, we measure the scattering from a ground glass, a 2-dimensional scatterer, as a function of the sample-$L_{1a}$ distance, for $14\mm \le d \le 23\mm$, in steps of 1 mm. The results are shown in Fig.~\ref{fig:glass}. Up to $q= 1\um^{-1}$ no significant differences are observed. For $1\um^{-1} \le q \le 4.6\um^{-1}$, the intensity level is systematically lower for $d =
14\mm$ and $d = 15\mm$, as compared to the other curves, while it is systematically higher for $d \ge 21\mm$, confirming the trend observed for the pinhole. For $d =
23\mm$ an anomalous increment in the
intensity profile is visible at $q = 4.6\um^{-1}$ and the same happens at higher $q$ vector for smaller
distances. This extra-signal is due to aberrations affecting rays scattered at high $q$
vectors and impinging close to the lens
edge. The ray-tracing software shows that these rays are systematically bent towards the optical axis, resulting in a spurious increase of the measured $I(q)$. At even larger $q$, the intensity decreases steeply because of
vignetting. To quantify the discrepancy between the ten intensity
distributions, the ratio between each curve and the intensity
profile measured for $d = 18\mm$ is shown in the inset of
Fig.~\ref{fig:glass}. If we consider the five distances between
$16\mm$ and $20\mm$, the ratio varies between 0.87 and 1.07 (horizontal black lines in the inset), and the maximum accessible scattering vector is $5\um^{-1}$. Thus, the response of the instrument for a $5\mm$ thick sample centred in $d =
18\mm$ is flat to within $\approx \pm 10\%$. If we consider the
two distances $18\mm$ and $19\mm$ (corresponding to a $2\mm$-thick sample) the
ratio deviates from 1 by less than $5\%$ and the maximum accessible $q$ vector is
$5.2\um^{-1}$.

\begin{figure}
\includegraphics[scale=1.8]{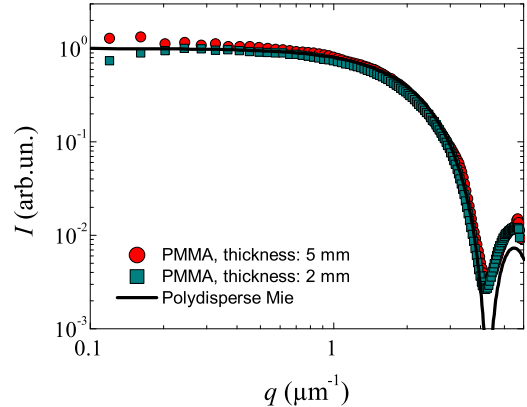}
\caption{\label{fig:pmma}Intensity distribution measured for a
sample of PMMA spheres with an average diameter of $2.09~\mathrm{\mu m}$
suspended in a mixture of decalin and tetralin and placed in two cells
of thickness $5\mm$ and $2\mm$. Both
datasets have been rescaled so that $I(q \rightarrow 0) = 1$.
The black line is the form factor calculated using the Mie theory.}
\end{figure}

We test the setup performances on a typical 3-dimensional sample by
measuring the scattering from a suspension of poly(methyl methacrylate)(PMMA) spheres with an average diameter of
$2.09~\mathrm{\mu m}$ and relative polydispersity of $2.5\%$, as obtained from
transmission electron microscopy (TEM). The particles are sterically stabilized by a thin layer of chemically grafted poly-
12-hydroxystearic acid (PHSA), and are suspended in a mixture of decalin and tetralin (0.676/0.324 v/v at $20 \degree
C$). The measurements have been performed at $26.5 \degree \mathrm{C}$; at this
temperature the solvent refractive index is 1.492 and the PMMA
refractive index is  $n_{\text{PMMA}} \approx
1.501$, as estimated by preparing samples in slightly different decalin/tetralin mixtures and by taking $n_{\text{PMMA}}$ as the refractive index of the solvent for which the scattered intensity is minimal. The
suspension was filtered through a $5\um$ PTFE membrane and then
placed in two cells of thickness $5\mm$ and
$2\mm$, respectively. In the first case, the sample volume fraction was $\varphi = 0.2\%$ while for the thinner cell it was $\varphi = 0.4\%$. Fig.~\ref{fig:pmma} shows the intensity
distributions measured for the two samples. Data have been
rescaled so that $I(q\rightarrow0)=1$. The line is the form factor from the Mie theory~\cite{Kerker} applied to polydisperse spheres, calculated using the MiePlot software~\cite{mieplot}, for the size distribution obtained by TEM. The agreement between the data and the theory is very good up to $4\um^{-1}$. At higher $q$, the intensity distribution is
systematically larger than the theoretical curve. This discrepancy is unlikely to stem mainly from a non-uniform angular response of the setup, since the tests with the pinhole and the ground glass suggest that deviations should compensate once integrated over the thickness of the sample. Other possible explanations include deviations from a perfectly spherical particle shape (as observed in the TEM images), scattering from the PHHS corona that may become non-negligible when matching closely the PMMA refractive index, and multiple scattering. Concerning the latter point, we note that although the suspension transmission was very high ($90\%$), even modest amounts of multiple scattering would tend to increase $I(q)$ in those angular ranges where the signal is markedly lower than for nearby $q$ vectors, as observed here.

\begin{figure}
\includegraphics[scale=1.8]{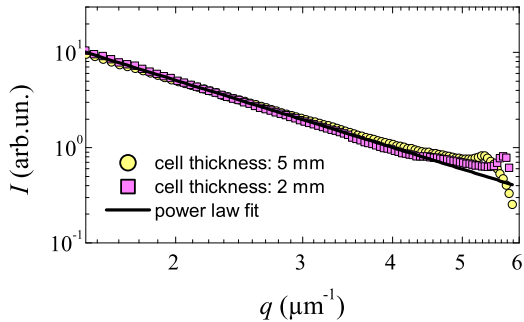}
\caption{\label{fig:aggreg} Intensity distributions measured for fractal
aggregates of latex spheres placed in cells of thickness $5\mm$ and $2\mm$. The plot zooms into the high $q$ range
$1 ~\mathrm{\mu m^{-1}} = q = 6  ~\mathrm{\mu m^{-1}}$.
The black curve is a power law with exponent $-d_{f} = -2.3$,
obtained by fitting the two experimental curves in the range
$1.5 ~\mathrm{\mu m^{-1}} \le q \le 3.5  ~\mathrm{\mu m^{-1}}$}
\end{figure}

To further test the angular response of the instrument for $q  \ge 4\um^{-1}$, we measure $I(q)$
for fractal aggregates made of colloidal particles. Fractal aggregates provide a convenient means to test the apparatus, since the scattering profile has a simple power-law dependence on scattering vector~\cite{M.CarpinetiF.FerriM.GiglioE.Paganini1990}: $I(q) \sim q^{-d_f}$, with $d_f$ the fractal dimension of the aggregate. While the fractal regime only extends beyond the particle size and below the aggregate dimension, it is easy to have the $I \sim q^{-d_f}$ regime covering the whole MALS range, by choosing appropriately the particle size and the aggregation time. To form the aggregates, we start by preparing a particle suspension of
latex spheres with a diameter of $19~\mathrm{nm}$ in a buoyancy matching mixture of Milli-Q water and $\mathrm{D_{2}O}$ (0.5/0.5 v/v), at $\varphi = 6 \times 10^{-4}$. Separately, a $30~\mathrm{mM}$ solution of $\text{MgCl}_{2}$ is prepared, using the same solvent. Equal amounts of the particle suspension and the salt solution are mixed directly in the scattering cell (cells with $L=5\mm$ and $L=2\mm$ were used), just before starting measurements that are run for about 17 hours. Initially, $I(q)$ exhibits a power law tail at large $q$ and a
rolloff at lower $q$ (data not shown), corresponding to the size of the aggregates. As clusters keep growing, the rolloff moves to smaller $q$, eventually exiting from the MALS window.
Figure~\ref{fig:aggreg} shows the intensity distribution
measured in the two scattering cells at later stages, focussing on the range $1\um^{-1} \le q \le 6\um^{-1}$. The black curve is a power law with exponent $-d_f$, obtained by fitting the
two experimental curves in the range $1.5 \um \le q \le 3.5\um^{-1}$, where the previous tests indicate that the instrument response is flat.
We find $d_f = 2.3$, consistent with typical values for colloidal aggregates~\cite{M.CarpinetiF.FerriM.GiglioE.Paganini1990}. The fit is in excellent agreement with the data below $q \le 3.5\um^{-1}$. The agreement remains quite good up to $q \approx 5\um^{-1}$, where a higher than expected intensity is measured, due to aberrations. At $q = 5\um^{-1}$ deviations are $15 \%$ and $24 \%$ for the $L=2\mm$ and $L=5\mm$ cells, respectively.

To summarize, the SLS tests indicate that the angular response of the setup is flat for thin samples ($L \le 2 \mm$) placed at $d=18\mm$, while for thicker samples $I(q)$ tends to be overestimated beyond $q \approx 4 \um^{-1}$. Even in the worst case that we have tested (a 5-mm thick cells), deviations do not exceed $24\%$ for $q$ vectors as large as $5\um^{-1}$.

\subsection{Dynamic Light Scattering}

\begin{figure}
\includegraphics[scale=1.8]{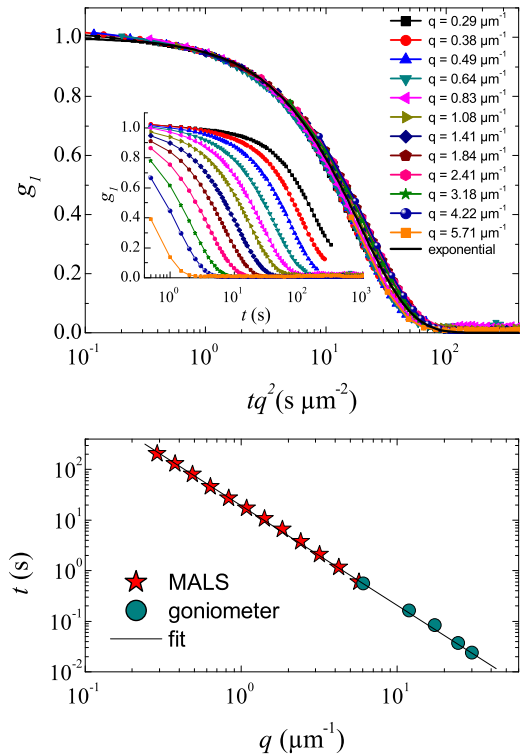}
\caption{\label{fig:dynamic} a) The twelve correlation
functions for a sample of polystyrene spheres with diameters of $530~\mathrm{nm}$
diluted in a mix of water/glycerol 0.3w/0.7w are shown as a function of time
rescaled by $q^{2}$. All data collapse onto a single master curve that is well fitted by
an exponential decay correlation (black line). The inset shows the twelve
original correlation functions on a semilogarithmic plot.
b) The decay time obtained by fitting
each correlation function to an exponential decay, as a
function of $q$, is shown in a double logarithmic scale (stars). The
circles are the decay time measured for the same sample
but at larger scattering angles using a conventional DLS apparatus.}
\end{figure}

To test the DLS capabilities of our apparatus, we measure the field time autocorrelation function of submicron Brownian particles, for which one expects $g_1(\tau) = \exp\left (-Dq^2\tau \right)$~\cite{Berne1976}, where $D$ is the particle diffusion coefficient. Twelve annuli of pixels were processed simultaneously in real time, at a rate of 4 images per second and using a single exposure time $t_{\text{exp}}=5~\text{msec}$. The annulus average radius ranged from $70 ~\mathrm{pixels}$ to $1253 ~\mathrm{pixels}$, logarithmically spaced
with a scaling factor of 1.3 between one annulus and the next one. The width of the innermost annulus was $1~\mathrm{pixel}$, corresponding to a relative spread in scattering vector $\delta q/q = 1.4\%$. The same ratio was kept for all other annuli. With this annuli selection, the time correlation function is measured in the range $0.29 \um^{-1} < q < 5.71 \um^{-1} $, corresponding to a variation of three orders of magnitude in the characteristic decay time of $g_1$. The sample is a dilute suspension of polystyrene spheres with diameter $530~\mathrm{nm}$ and polydispersity $5\%$ (nominal values), suspended in a mixture of
Milli-Q water/glycerol (approximately 0.3/0.7 w/w, $n = 1.421$), at $\varphi =
6.4 \times 10^{-5}$ and loaded in a cell with $L=5\mm$. The particle size and the solvent viscosity were chosen so that the decay time of $g_1$ be large enough to allow for real time measurements. Data were collected over 30 minutes, 9 hours after loading the cell.

The inset of Fig.~\ref{fig:dynamic} a) shows the twelve correlation functions on a semilogarithmic plot. An offset has been subtracted from the raw data, since the baseline is typically somehow larger than zero, due to contributions from stray light that are not fully correct for by the background subtraction described in Ref.~\cite{Cipelletti1999}. This offset is at most $0.15$ for the smallest and largest $q$ and $ \lesssim 0.05$ for most $q$ vectors. In the main figure, the same data are plotted as a function of time rescaled by $q^2$. All data collapse onto a single master curve that is well fitted by an exponential decay (line in Fig.~\ref{fig:dynamic} a). Both the data collapse and the exponential shape are in agreement with what expected for particles undergoing Brownian diffusion. Note that due to the $q^2$ scaling, the quality of the collapse is quite sensitive to any error in the determination of the scattering vector. Thus, the data of Fig.~\ref{fig:dynamic}a) confirm the validity of the $q$ calibration. Figure~\ref{fig:dynamic}b) shows the decay time obtained by fitting separately each correlation function to an exponential decay, as a function of $q$, in a double logarithmic scale (stars). The circles are the decay time measured for the same sample but at larger scattering angles, $20\dg \le \theta \le 120\dg$ ($6 \um^{-1} < q < 30 \um^{-1} $), using a conventional DLS apparatus (Amtec goniometer equipped with a Brookhaven BT9000 correlator and an Argon
ion laser operating at $\lambda = 514 ~\mathrm{nm}$). The DLS and MALS data follow the same trend, showing that not only does the CCD-based apparatus capture well the shape and $q$ dependence of the correlation functions, but it also measures correctly their decay time on an absolute scale. The line is a power law fit to the data, with a fixed exponent $-2$ as inferred from diffusive motion, showing overall a very good agreement with both data sets. The first four points measured by the MALS apparatus were not included in the fit, because they slightly deviate from the $q^{-2}$ trend. This slight discrepancy is most likely due to additional ballistic dynamics that superimpose to the Brownian motion and originate from convective motion. Indeed, we find this effect to be even more relevant right after filling the cell, when temperature fluctuations and gradients are not stabilized yet. Such ballistic motion, while negligible with respect to diffusion on the small length scales probed by traditional DLS, become increasingly important as $q$ is decreased, since for ballistic motion the relaxation time of $g_1$ scales as $q^{-1}$, as opposed to $q^{-2}$ for Brownian diffusion~\cite{Berne1976}. Therefore, their impact becomes eventually visible, as observed here in the low $q$ spectrum of the MALS apparatus.

\section{Conclusions}
\label{sec:conclusions}

We have presented a static and dynamic light scattering apparatus that covers the angular range intermediate between those typically covered by small-angle and wide angle setups. Thanks to a novel optical layout, our apparatus does not require custom-made optical or opto-mechanical components, which should make it relatively easy to duplicate it even by groups with little experience in optics.

As a final remark, we note that we have also tested an alternative choice of lenses, while keeping the same overall layout. In this alternative scheme, the first two pairs of lenses are replaced by single aspherical lenses with a large numerical aperture, while the third pair of lenses is replaced by a single biconvex spherical lens. Details of the alternative lenses are provided in the Supplementary Information. We find that the performances at the largest angles are comparable, while the flare at small angles is slightly larger when using the aspherical lenses, presumably because the optical quality of their surfaces is not as good as for the spherical lenses.

\begin{acknowledgments}
We thank J. Barbat for designing and machining the sample holder and A. Schofield for providing us with the PMMA particles. We gratefully acknowledge discussions with J. Oberdisse and L. Ramos and funding from the CNRS and the ANR (Contract No. ANR-09-BLAN-0198, COMET).
\end{acknowledgments}


\end{document}